# Bibliometric Mapping of AI-Supported Social Presence in Online Learning Environments: Trends, Collaboration, and Thematic Directions

Almer Balingit Gamboa
College of Education
Pampanga State University
Mexico, Pampanga, Philippines
abgamboa@pampangastateu.edu.ph

Erika Mamucud Pineda
College of Education
Pampanga State University
Bacolor, Pampanga, Philippines
empineda@pampangastateu.edu.ph

Rhiziel Peña Manalese
College of Arts and Sciences
Pampanga State University
Bacolor, Pampanga, Philippines
rpmanalese@pampangastateu.edu.ph

Aileen Pobre De Leon
College of Computing Studies
Pampanga State University
Bacolor, Pampanga, Philippines
apdeleon@pampangastateu.edu.ph

Vernon Grace Magat Maniago
College of Computing Studies
Pampanga State University
Sto. Tomas, Pampanga, Philippines
vgmmaniago@pampangastateu.edu.ph

Jan Henry Beltran Sunga
College of Education
Pampanga State University
Bacolor, Pampanga, Philippines
jhbsunga@pampangastateu.edu.ph

Agnes Romero Regala
College of Education
Pampanga State University
Bacolor, Pampanga, Philippines
arregala@pampangastateu.edu.ph

Roque Francis Badajos Dianelo
College of Education
Pampanga State University
Mexico, Pampanga, Philippines
rfbdianelo@pampangastateu.edu.ph

John Paul Palo Miranda*
College of Computing Studies
Pampanga State University
Mexico, Pampanga, Philippines
jppmiranda@pampangastateu.edu.ph

## Abstract

This study examines the development, influence, and collaboration patterns in AI-supported social presence research within online learning environments. Utilizing 59 open-access empirical studies from Scopus, the study applies citation analysis, co-authorship mapping, institutional analysis, and keyword clustering using Python-based bibliometric tools. Findings reveal an upward trend in publications since 2020, with research focusing on engagement, AI tools, instructional design, and ethical issues. While countries such as the United States and Brazil are leading contributors, international collaboration remains limited. Ethical concerns related to trust and fairness are emerging but underexplored. The study highlights the importance of ethical integration, interdisciplinary collaboration, and learner-centered AI applications in education.

*Correspondence author

**ACM Reference Format:**
Almer Balingit Gamboa, Erika Mamucud Pineda, Rhiziel Peña Manalese, Aileen Pobre De Leon, Vernon Grace Magat Maniago, Jan Henry Beltran Sunga, Agnes Romero Regala, Roque Francis Badajos Dianelo, and John Paul Palo Miranda. 2025. Bibliometric Mapping of AI-Supported Social Presence in Online Learning Environments: Trends, Collaboration, and Thematic Directions. In *2025 the 9th International Conference on Education and Multimedia Technology (ICEMT) (ICEMT 2025), July 29–August 01, 2025, Osaka, Japan.* ACM, New York, NY, USA, 5 pages. https://doi.org/10.1145/3761843.3761902

## 1 INTRODUCTION

Social presence refers to the sense of interpersonal connection and emotional engagement that learners experience in digital learning environments [1, 6, 13]. It is essential in online education because it supports motivation, satisfaction, and learning outcomes [9]. The Community of Inquiry framework identifies social presence as a foundational element of meaningful virtual learning [13]. However, creating social presence remains difficult in online settings where face-to-face interaction is absent [1]. Negative academic self-concepts and reduced engagement in fully online courses have been linked to issues in interaction, pedagogy, technology, and personal learning environments [2]. These challenges highlight the limitations of current online systems in supporting learner presence and confidence. Similar problems have also been observed in metaverse environments, where elements such as immersion, embodiment, and copresence significantly influence the development of social relationships [4]. These findings reinforce the need to design systems that strengthen interpersonal connection and social engagement in virtual contexts.

Recent advancements in artificial intelligence (AI) offer potential solutions through the use of chatbots, intelligent agents, adaptive systems, and emotion-aware interfaces [11, 14, 15]. These technologies aim to simulate human interaction and provide tailored support, but most studies focus only on isolated aspects such as engagement or feedback. Broader investigations that examine how AI systematically supports social presence are still limited. For this reason, this study uses a bibliometric review to examine the development, collaboration, and thematic directions of this emerging field [3, 5]. It identifies publication trends, research networks, and conceptual clusters. This study specifically analyzes open-access empirical studies indexed in Scopus and aims to: (1) examine annual publication trends and geographical distribution; (2) identify leading authors, sources, and institutions; (3) map collaboration networks and assess the extent of interdisciplinary and international engagement; (4) determine dominant research themes through keyword and thematic clustering; and (5) identify knowledge gaps.

## 2 METHOD

This study used bibliometric analysis to investigate trends, impact, author collaboration, and conceptual structures in empirical research on AI-driven social presence in learning. The analysis focused on open-access publications indexed in the Scopus database, which offers comprehensive coverage of peer-reviewed literature. The dataset included 72 initial documents through an advanced search conducted in February 2025. The search query applied multiple TITLE-ABS-KEY terms to capture relevant studies. Combined phrases include "social presence," "AI-driven learning," "artificial intelligence in education," "AI tutors," "chatbots in education," "virtual tutors," "AI-mediated learning," and "intelligent tutoring systems" with terms related to learning modes like "online learning," "e-learning," "distance education," and "digital learning." Additional terms included "student engagement," "human-AI interaction," "learner perception," "trust in AI," and "educational technology." Non-empirical articles, duplicates, and irrelevant studies were excluded to ensure relevance and quality. The final dataset includes 59 empirical studies.

The analysis used citation metrics, co-authorship networks, institutional affiliations, and keyword clustering to examine the structure and evolution of the field. Citation and authorship patterns were analyzed using *NetworkX*, while keyword clustering was performed with *scikit-learn* libraries. The study assessed annual publication trends, leading contributors, and dominant research themes. Data processing and visualization were conducted using Python libraries, including *pandas* for data management and *NetworkX* for network mapping. The use of bar charts, line graphs, and co-occurrence networks supported a comprehensive understanding of the field's development and emerging areas of interest.

## 3 RESULTS AND DISCUSSION
### 3.1 Publication Frequency

The number of studies on AI-driven social presence in learning has increased steadily, with a sharp rise beginning in 2020. Minimal research activity occurred prior to 2010, with only one publication each in 2006, 2009, and 2010. From 2013 to 2019, output ranged between 2 and 4 publications per year, suggesting gradual growth. A shift occurred in 2020, with research activity intensifying in response to the growing reliance on online education. The number of studies rose to 5 in 2021, 10 in 2022, 9 in 2023, and peaked at 13 in 2024. Alongside this increase, authorship trends show broader scholarly participation, with over 30 unique contributors each in 2022 and 2024, compared to fewer than 10 before 2020. These patterns suggest growing academic attention and broader disciplinary engagement with AI-enhanced social presence.

### 3.2 Leading Publishers

Among the 47 unique sources, five publishers had the highest number of studies in social presence in AI-driven learning (Table 1). Sustainability (Switzerland) leads with 3 publications, while Journal of Information Technology Education: Research, Computers and Education, Frontiers in Psychology, and International Review of Research in Open and Distributed Learning each have 2 publications. The average citation count per journal varies significantly, with International Review of Research in Open and Distributed Learning showing the highest average citation impact (119.5 citations per paper), followed by Computers and Education (46.0 citations per paper). This variation suggests that while some journals contribute frequent publications, others publish fewer but highly influential studies. The concentration of research in a limited number of journals indicated that while social presence in AI-driven learning is growing, dissemination remains somewhat concentrated in specific educational and interdisciplinary technology journals.

### 3.3 Author Productivity and Collaboration Networks

The dataset includes 191 unique authors contributing to 59 empirical studies on AI-driven social presence in learning. Author productivity is low, with each author contributing to only one publication indicating that the field remains in its early stages of development and driven by small, project-specific collaborations rather than long-term research programs. A few individuals demonstrate strong collaborative involvement. Authors such as Suemoto, Tomás, Andretta, and Vidor each maintain 17 co-authorship connections which makes them the most central figures within the largest co-authorship cluster. While these authors represent concentrated collaboration within specific groups, the broader co-authorship network remains fragmented with minimal links between clusters.

The top five co-authorship clusters highlight this fragmentation but also demonstrate thematic diversity (Figure 1). Cluster 1 forms the most interconnected group which is from the same institution that focuses on AI applications in learning with strong internal collaboration. Cluster 2 consists of multi-institutional researchers working on adaptive systems and human-AI interaction which indicates broader disciplinary integration. Cluster 3 centers on data-driven learning analytics and predictive modeling, with authors contributing to scalable AI implementations. Cluster 4 represents a regional team from the Gulf region and works on culturally specific applications of educational AI. Cluster 5 includes Eastern European researchers engaged in AI-supported personalization and assessment that is linked to national or cross-European education initiatives. These clusters reflect the emergence of research communities with shared interests, though collaborative linkages across clusters remain limited.

Table 1: Top publishers

| Rank | Source Title | Number of Publications |
| --- | --- | --- |
| 1 | Sustainability (Switzerland) | 3 |
| 2 | Journal of Information Technology Education: Research | 2 |
| 3 | Computers and Education | 2 |
| 4 | Frontiers in Psychology | 2 |
| 5 | International Review of Research in Open and Distributed Learning | 2 |
| 6-10 | Other Sources (Condensed) | 1 each |

Table 2: Thematic clusters

| Cluster | Top Keywords | Label |
| --- | --- | --- |
| 1 | blended learning, student engagement, engagement, covid-19, online learning | Engagement and Online Learning Contexts |
| 2 | social presence, teaching presence, higher education | AI Tools and Social Presence |
| 3 | distance education, community of inquiry, instructional strategies | Instructional Design and Pedagogical Frameworks |
| 4 | intelligent tutoring systems, architectures for educational technology system | Emerging Technologies and Ethical Considerations |

## 3.4 Institutional and Geographical Distribution

The contributing authors are affiliated with institutions from a wide range of countries. The most frequently represented institutions include Abdelmalek Essaadi University (Morocco), Unitec Institute of Technology (New Zealand), University of Tasmania (Australia), and Stockholm University (Sweden). While institutional contributions are diverse, no single institution dominates the field. At the national level, the United States is the leading contributor with 26 documents, followed by Brazil (12), Canada (9), Australia (8), and Indonesia (6). This distribution shows that interest in AI-driven social presence in learning spans both Western and non-Western contexts. However, most collaborations appear to occur within the same institution or country, with limited evidence of sustained international partnerships. Expanding cross-border research collaboration may improve knowledge integration and contribute to a more globally connected body of work in this field.

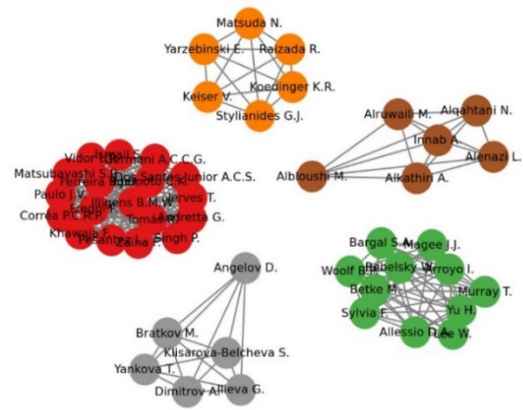

Figure 1: Top co-authorship network

## 3.5 Keyword Network, Thematic Clustering, Evolving Trends

The keyword co-occurrence network illustrates how this field is conceptually structured (Figure 2), with frequently occurring and interconnected terms such as "online learning," "student engagement," "social presence," "instructional strategies," and "community of inquiry." These terms form the core of the field and reflect dominant interests in AI-supported interaction, motivation, and pedagogy. The network displays two main thematic orientations: one that links "community of inquiry" with "instructional strategies" and "distance education," and another that connects "intelligent tutoring systems" with "architectures for educational technology." Thematic clustering analysis identified four major areas of research (Table 2). Cluster 1, Engagement and Online Learning Contexts, includes studies on how remote and blended learning environments affect engagement, particularly during COVID-19. Cluster 2, AI Tools and Social Presence, focuses on AI applications such as chatbots and intelligent systems designed to enhance interaction. Cluster 3, Instructional Design and Pedagogical Frameworks, highlights research grounded in teaching strategies and models like the Community of Inquiry. Cluster 4, Emerging Technologies and Ethical Considerations, includes topics such as intelligent tutoring systems and ethical implications of AI. These results suggests that the field has developed strong conceptual foundations in instructional and technical areas, while ethical considerations and cross-disciplinary integration remain limited.

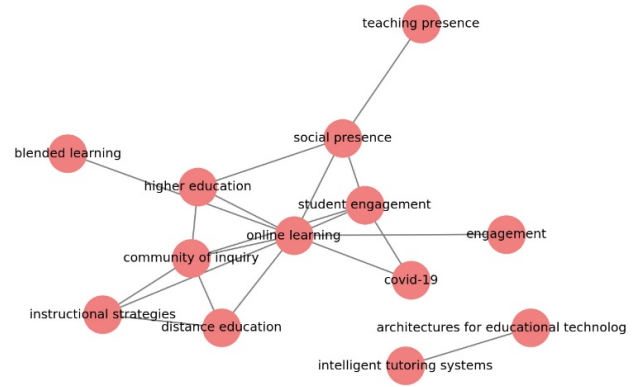

Figure 2: Keywords co-occurrence network

Table 3: Top 5 most cited documents

| Title | Authors | Year | Citations |
| --- | --- | --- | --- |
| Learners' perspectives on what is missing from online learning: Interpretations through the community of inquiry framework | Stodel et al. | 2006 | 175 |
| Supersizing e-learning: What a CoI survey reveals about teaching presence in a large online class | Nagel & Kotzé | 2010 | 113 |
| Prediction and Localization of Student Engagement in the Wild | Kaur et al. | 2019 | 100 |
| A framework of implementing strategies for active student engagement in remote/online teaching and learning during the covid-19 pandemic | Ahshan | 2021 | 75 |
| Teaching music online: Changing pedagogical approach when moving to the online environment | Johnson | 2017 | 73 |

### 3.6 Citation Analysis

The top five most cited studies in this field examined learner engagement, the Community of Inquiry framework, and AI-driven feedback (Table 3). Stodel et al. [12], with 175 citations, investigated learners' views on what is lacking in online education. The study used the Community of Inquiry framework and identified five missing elements: robust dialogue, spontaneity, mutual recognition, personal connection, and the development of online learning skills. Nagel and Kotzé [10], cited 113 times, applied the same framework to a large postgraduate class and demonstrated that creative use of digital tools such as peer review can enhance teaching, cognitive, and social presence in large-scale online environments. Kaur et al. [8], with 100 citations, introduced a dataset that used behavioral cues and deep learning to predict and localize student engagement in online video content. This study contributed to the advancement of AI-supported learning systems such as MOOCs and intelligent tutoring systems. Ahshan (2021), with 75 citations, proposed a framework that combined pedagogy, educational technologies, and e-learning platforms to improve student engagement during the COVID-19 pandemic. Johnson [7], cited 73 times, explored how university music instructors shifted their pedagogy when teaching online and found that they adopted collaborative and constructivist strategies aligned with the Community of Inquiry model. These studies highlight the centrality of student interaction, technological integration, and pedagogical change. The citation rates range from 1.30 to 3.13 per year which showed a consistent scholarly influence in this field.

## 4 CONCLUSION, RECOMMENDATIONS, and limitations

This study provides a bibliometric overview of open-access empirical research on AI-driven social presence. It shows that this area has emerged as a distinct field within educational technology. Although interest has increased since 2020, the findings indicate that current research is conceptually fragmented, methodologically narrow, and limited in terms of cross-institutional collaboration. Thematic mapping shows recurring attention to engagement and instructional design, but ethical concerns and theoretical integration are still weak. Future research should focus on sustained inquiry that deepens conceptual models, evaluates AI applications in different learning contexts, and investigates long-term effects on learner interaction. Researchers should also develop frameworks that clearly connect AI functionality with social presence outcomes in ways that support meaningful pedagogy. This study has limitations. It only includes open-access empirical studies indexed in the Scopus database. As a result, it may have excluded important research from other sources. Future reviews should use additional databases such as Web of Science, Google Scholar, and ERIC and include non-open-access studies to present a more complete and representative view of the field.

## 5 PRACTICAL IMPLICATIONS

This study indicated that research on AI-driven social presence remains limited in scope and uneven in distribution across regions and institutions. Although interest in this area has grown since 2020, the field lacks sustained collaboration, comprehensive conceptual development, and integration of ethical concerns such as trust and fairness. Future research must address these gaps by exploring how AI can meaningfully support social presence across diverse educational contexts. There is a need for studies that examine the effectiveness of specific AI tools, evaluate learner experiences, and develop theoretical models that link social presence with AI-mediated interaction. Expanding empirical work in these directions will strengthen the field and support the design of socially responsive and inclusive digital learning environments.